\documentclass[%
reprint,
superscriptaddress,
amsmath,amssymb,floatfix,
aps]{revtex4-2}
\usepackage{graphicx}% Include figure files
\usepackage{dcolumn}% Align table columns on decimal point
\usepackage{bm}% bold math
\usepackage{gnuplottex}
\usepackage{epstopdf}
\usepackage{float}
\usepackage[normalem]{ulem}
\usepackage{epstopdf}
\usepackage[dvipsnames]{xcolor}
\usepackage[%  
    colorlinks = true,
    linkcolor = blue,
    urlcolor  = blue,
    citecolor = blue,
    anchorcolor = blue
]{hyperref}

\newcommand{\mnau}{Mn\textsubscript{2}Au }
\newcommand{\neel}{N\'eel }

\begin{document}
	
\title{Multiscale Modelling of the Antiferromagnet Mn\textsubscript{2}Au: From ab-initio to Micromagnetics}

\author{Joel Hirst}
\affiliation{\mbox{Materials \& Engineering Research Institute, Sheffield Hallam University, Howard Street, Sheffield S1 1WB, United Kingdom}}
\author{Unai Atxitia}
\affiliation{Dahlem Center for Complex Quantum Systems and Fachbereich Physik, Freie Universitat Berlin, 14195 Berlin, Germany}
\affiliation{Instituto de Ciencia de Materiales de Madrid, CSIC, Cantoblanco, 28049 Madrid, Spain}
\author{Sergiu Ruta}
\affiliation{\mbox{Materials \& Engineering Research Institute, Sheffield Hallam University, Howard Street, Sheffield S1 1WB, United Kingdom}}
\author{Jerome Jackson}
\affiliation{Scientific Computing Department, STFC Daresbury Laboratory, Warrington WA4 4AD,United Kingdom}
\author{Leon Petit}
\affiliation{Scientific Computing Department, STFC Daresbury Laboratory, Warrington WA4 4AD,United Kingdom}
\author{Thomas Ostler}
\affiliation{Department of Physics \& Mathematics, University of Hull, Hull HU6 7RX, UK}
\affiliation{\mbox{Materials \& Engineering Research Institute, Sheffield Hallam University, Howard Street, Sheffield S1 1WB, United Kingdom}}
\affiliation{\mbox{Department of Engineering and Mathematics, Sheffield Hallam University, Howard Street, Sheffield S1 1WB, United Kingdom}}
\date{\today}

\begin{abstract}
	Antiferromagnets (AFMs) are strong candidates for the future spintronic and memory applications largely because of their inherently fast dynamics and lack of stray fields, with \mnau being one of the most promising. For the numerical modelling of magnetic material properties, it is common to use \textit{ab-initio} methods, atomistic models and micromagnetics. However, each method alone describes the physics within certain limits. Multiscale methods bridging the gap between these three approaches  have been already proposed for ferromagnetic materials.
	Here, we present a complete multiscale model of the AFM \mnau as an exemplar material, starting with results from ab-initio methods going via atomistic spin dynamics (ASD) to an AFM Landau-Lifshitz-Bloch (AFM-LLB) model. Firstly, bulk \mnau is modelled using a classical spin Hamiltonian constructed based on earlier first-principles calculations. Secondly, this spin model is used in the stochastic Landau-Lifshitz-Gilbert (LLG) to calculate temperature-dependent equilibrium properties, such as magnetization and magnetic susceptibilities. Thirdly, the temperature dependent micromagnetic parameters are used in the AFM-LLB. We validate our approach by comparing the ASD and AFM-LLB models for three paradigmatic cases; (i) Damped magnetic oscillations, (ii) magnetization dynamics following a heat pulse resembling pump-probe experiments, (iii) magnetic domain wall motion under thermal gradients.
\end{abstract}

\maketitle

\section{\label{intro} Introduction}
The interest in antiferromagnetic (AFM) materials has increased in recent years for several reasons. From fundamental questions regarding the effects of atomic-scale spin interactions on the macroscale magnetic configurations \cite{dzyaloshinskii1964theory, chakravarty1988low, manousakis1991spin, doi:10.1126/sciadv.abn3535}, to contenders for driving future technologies of smaller, faster, and more energy efficient devices \cite{PhysRevLett.113.157201, Marti2014, Legrand2020, Nemec2018, zelezny2018, PhysRevLett.126.187602}.
Possible applications are in the fields of terahertz (THz) science, ultrafast spin dynamics and spin-caloritronics -- combined spin transport and heat.
\\
\\
AFMs are robust against strong external magnetic fields and have intrinsically fast THz magnetization dynamics compared to the (gigahertz) GHz dynamics found in ferromagnets. It has been known since the late 1940's that the resonant modes of certain AFMs can reach the THz range \cite{kittel1, kittel2}. This frequency enhancement in AFMs emerges at the macroscale from the interplay of the magnetic anisotropy and the atomic-scale antiferromagnetic spin exchange interaction. If antiferromagnets are to be used in THz emitters and spintronic devices \cite{Checinski2017, PhysRevLett.116.207603, Khymyn2017, Kampfrath2011}, it becomes essential to accurately calculate, from first-principles, the nature and values of the anisotropy and exchange interactions between spins. Microscopic interactions also play a key role in the field of ultrafast spin dynamics \cite{Beaurepaire}. Experiments have demonstrated an ultrafast and more energy efficient magnetic order quenching in AFM than in FMs \cite{PhysRevLett.119.197202}. Further understanding and control of the ultrafast spin dynamics in AFMs requires not only accurate determination of the exchange interactions, but also how the magnetic system reacts to an ultrafast load of heat from femtosecond laser pulses.
\\
\\
Metallic AFMs are convenient for all of those aforementioned applications. For instance, the antiferromagnets CuMnAs and Mn$_2$Au have been proposed \cite{PhysRevApplied.9.064040, Chen2021, PhysRevApplied.16.014037, MACA20121606, Olejnik2017, PhysRevLett.118.057701} for several device applications because their magnetic state can be controlled via electric currents, and they can be read out via their magnetoresistive properties \cite{Bodnar2018,Bommanaboyena2021}. They have relatively high critical temperature with experimental measurements placing the \neel temperature of \mnau between 1300 and 1600 K \cite{Barthem2013} and around 480 K for CuMnAs \cite{Wadley2015}, making them ideal for memory and spintronics applications. While initial studies focused on CuMnAs rather than Mn$_2$Au, it has been demonstrated that the Joule heating created by electric currents puts the system too close to the critical temperature, so that undesired thermal fluctuations strongly influence the dynamics. In this regard, the much higher critical temperature of \mnau is preferred.
\\
\\
Established approaches to calculate the magnetic properties  include \textit{ab-initio} methods for the electronic structure at the atomic scale and generally at zero temperature \cite{vasp, siesta,espresso, sprkkr1, sprkkr2}, atomistic spin models coupled to a heat-bath for the calculation of thermal properties at the  \cite{Evans2014,upasd,spirit}, and micromagnetics for the calculation of the magnetic distribution at the micrometer scale \cite{OOMMF,MUMAX,BORIS}. \textit{Ab-initio} methods and atomistic spin models naturally describe AFMs. Modeling AFMs with micromagnetism is more problematic since the micromagnetic models are based on small variations of the magnetization at the nanoscale, while in AFMs  the spin orientation varies strongly at the atomic scale.
While micromagnetic modelling of AFMs remains a challenge, ferromagnetic models based on the Landau-Lifshitz-Gilbert (LLG) equation are now well established and used extensively in fundamental and device research. Several codes are available with a plethora of features and are implemented on various high performance computing platforms \cite{OOMMF,MAGPAR,MUMAX}. These micromagnetic models are particularly useful when attempting to model scenarios in which the dynamics occur across $\mu$m distances and where high frequency spinwaves do not contribute to the dynamics, for example, when modelling domain structures~\cite{Wohlhuter2015} or vortex core dynamics \cite{Moriya2008}.
One of the main drawbacks of conventional micromagnetic modeling is that the simulations remain at fixed temperature, rendering LLG based micromagnetics obsolete in scenarios where the temperature changes dynamically or when it varies spatially as transient changes in the macrosopin magnetisation are not naturally taken into account. 
\\
\\
It is possible to use atomistic spin dynamics (ASD) in scenarios where thermal effects play an important role. In ASD, each magnetic lattice site has a magnetic moment, the dynamics of which are simulated with the stochastic LLG equation \cite{LLG1, LLG2}. The magnetic atoms interacts with neighbouring moments which can be written as an extended Heisenberg Hamiltonian \cite{Evans2014, PhysRevB.83.024401}. Atomistic modelling has become an essential tool for modelling ultrafast magnetization dynamics \cite{Ostler2012, Radu2011} and temperature dependent effects such as laser pump-probe experiments \cite{PhysRevB.82.054415, doi:10.1063/1.3515928, PhysRevApplied.14.014077}. However, simulations of systems on $\mu$m lengthscales and $\mu$s timescales (such as the magnetic grains found in novel magnetic recording media) become computationally expensive, and a finite temperature micromagnetic approach valid for AFMs is needed. Along this line, extensive work already exists using the so-called Landau-Lifshitz-Bloch (LLB) finite temperature micromagnetic formalism to model ferro and ferrimagnets, including for FePt \cite{Kazantseva2008, Ostler2014}, GdFeCo \cite{Atxitia2012} and permalloy \cite{Hinzke2015}. The LLB framework has been used to model magnetic systems on device-level length and timescales. However, the LLB formalism for AFMs is incomplete. First attempts have been made to calculate the domain wall velocity due to a thermal gradient for a generic AFM using an LLB model \cite{chen_afm_llb}. However, the proposed model for AFM-LLB lacks a fundamental aspect of the magnetic properties at finite temperatures, the relaxation of the magnetization length. The correct description of the magnetization quenching dynamics (aka longitudinal relaxation dynamics) is essential for the modelling of ultrafast magnetization dynamics and switching. A revised version of the AFM-LLB that includes the recently derived exchange enhanced longitudinal effective damping \cite{florian} is necessary in order to extend the LLB framework to AFMs.
\\
\\
Here we present a complete multiscale model of the AFM Mn\textsubscript{2}Au. The model starts with parameters from Density Functional Theory (DFT) calculations \cite{sergiu}, then going via atomistic spin dynamics to a newly developed LLB equation for AFMs. We calculate the thermal equilibrium properties using ASD and use the results as input into an AFM-LLB model, where instead of having a spin at each atomistic site we simulate a collection of spins, known as a macrospin (in much the same way as is done in traditional LLG based micromagnetics). This approach, whereby one starts with results from first-principle calculations and end up with a description on micromagnetic length scales, is known as multiscale modelling, and this is the first of its kind for AFMs where both the transverse and longitudinal processes have been accurately within an LLB framework.
\\
\\
The work is broken down in the following way. We begin with an introduction of the atomistic spin model with the DFT parameters that are used to parameterise the atomistic Hamiltonian. We then present the transverse terms of the AFM LLB model and present results for the Antiferromagnetic Resonance (AFMR) and compare to ASD and analytical expressions. We then introduce the longitudinal term in the LLB model and compare the dynamics from ASD and LLB simulations for step changes in temperature as well as for ultrafast laser heating using a two-temperature model. Finally, as an example where both the longitudinal and transverse components of the AFM-LLB are important (and where ASD simulations take orders of magnitude more computational time than an LLB approach) we present a comparison between LLB, ASD and analytical expressions for the the domain wall (DW) motion due to a thermal gradient. 

\section{Atomistic Model}
\label{sec:atomistic}
	The dynamics of each atomistic spin is governed by the stochastic Landau-Lifshitz-Gilbert equation (s-LLG) equation:
	\begin{equation}
	\frac{\partial \mathbf{S}_{i}}{\partial t}=-\frac{\gamma_i}{\left(1+\lambda_i^{2}\right)\mu_i}\left[\mathbf{S}_{i} \times \mathbf{H}_{i}+\lambda_i \mathbf{S}_{i} \times\left(\mathbf{S}_{i} \times \mathbf{H}_{i}\right)\right]
	 \label{eq:sLLG}
	\end{equation}
	where $\mathbf{S}_i$ is a normalised unit vector of the spin at site $i$, $\lambda_i$ is the effective damping parameter, $\gamma_i$ is the
	gyromagnetic ratio, $\mu_i$ is the atomic magnetic moment and $\mathbf{H}_i$ is the effective field acting on the spin
	at site $i$. The effective field is the sum of a field-like stochastic term  and the negative derivative of the spin Hamiltonian:
	\begin{equation}
	\label{eq:eff_field}
	\mathbf{H}_{i}= \boldsymbol{\zeta}_{i}(t)-\frac{\partial \mathcal{H}}{\partial \mathbf{S}_{i}}
	\end{equation}
	where the Hamiltonian given by:
	\begin{equation}
	\mathcal{H}= -\sum_{\langle i j\rangle} J_{ij}\mathbf{S}_{i} \cdot \mathbf{S}_{j}-d_\eta \sum_{i}\left(S_{i}^{\eta}\right)^{2}-\mu_{s} \mathbf{B} \cdot \sum_{i} \mathbf{S}_{i}
	\end{equation}
	where $J_{ij}$ is the exchange interaction between site $i$ and site $j$, $\mathbf{B}$ is the Zeeman field and $d_\eta$ is the uniaxial anisotropy constant with $\eta = x,y,z$.
	The stochastic term $\boldsymbol{\zeta}_{i}(t)$ in Eq. \eqref{eq:eff_field} describes the coupling of the spin system to an external thermal bath and accounts for the thermal fluctuations. The noise processes are governed by the equations:
\begin{equation}
\begin{gathered}
\left\langle\zeta_{i}^{a}(t) \zeta_{j}^{b}(t)\right\rangle=2 \delta_{i j} \delta_{a b} \delta\left(t-t^{\prime}\right) \frac{\mu_{i} \lambda_{i} k_{B} T}{\gamma_{i}} \\
\left\langle\zeta_{i}^{a}(t)\right\rangle=0
\end{gathered}
\end{equation}
Where $a$ and $b$ are the Cartesian components and $T$ is the temperature of the system. The integration of the s-LLG equation is completed using the Heun integration method. The simulations are performed on an in-house code capable of running on graphical processing units (GPUs).
\\
\\
\begin{figure}[t]
\centering
\includegraphics[scale=0.25]{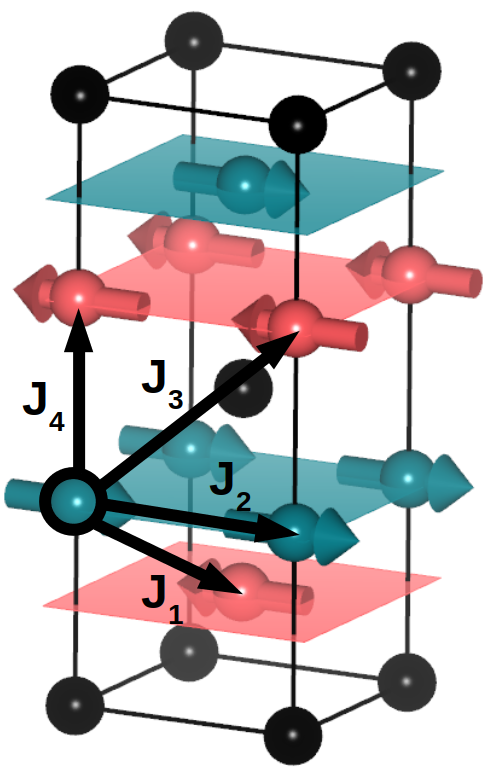}
\caption{The \mnau unit cell. The exchange interactions used in this work are labelled $J_1$ to $J_4$. The values can be found in Table \ref{tab:parameters} 
}
\label{fig:mnau_cell}
\end{figure}
\begin{table}[t!]
\begin{tabular}{|l|r|l|}
\hline
Constant                                              & \multicolumn{1}{l|}{Value}  & Unit    \\ \hline
Lattice constant, $a$                                 & 3.330                       & $\text{\normalfont\AA}$   \\
Lattice constant, $c$                                 & 8.537                       & $\text{\normalfont\AA}$   \\
Magnetic Moment, $\mu_s$                               & 3.8663                      & $\mu_B$ \\
Exchange constant, $J_1$                              & -5.3422                     & mRy     \\
Exchange constant, $J_2$                              & 0.6484                      & mRy     \\
Exchange constant, $J_3$                              & -0.6341                     & mRy     \\
Exchange constant, $J_4$                              & -6.8986                     & mRy     \\
Sum of FM exchange, $J_{0, \nu \nu}$               & 2.5934                      & mRy     \\
Sum of AFM exchange, $J_{0, \nu \kappa}$           & -30.8040                     & mRy     \\
Uniaxial Anisotropy in $z-$direction, $d_{||}$          & -0.0090                     & mRy     \\
Uniaxial Strain Anisotropy in $z-$direction, $d^*_{||}$ & 0.0004                      & mRy     \\ \hline
\end{tabular}
\caption{Parameters used in the Hamiltonian for the atomistic modelling of Mn$_2$Au.}
\label{tab:parameters}
\end{table}
For the atomistic simulations of Mn\textsubscript{2}Au, we take parameters for the exchange constants, magnetic moment and anisotropy from previous work by Ruta et al \cite{sergiu}. 
Using the LMTO-ASA method \cite{PhysRevB.12.3060}, they calculate a strong negative uniaxial anisotropy constant along the $z-$direction meaning the \neel vector has no preferential direction in the plane.
We opt to use an additional uniaxial anisotropy term along the [100] direction induced by in-plane strain with a value of 0.0004 mRyd/Mn as calculated by Shick et al \cite{shick_mn2au}. In such a way, the system has a single favoured orientation of the \neel vector. We use the first first four nearest neighbour interactions between Mn sites as illustrated
in Fig. \ref{fig:mnau_cell}. The  parameters used in the atomistic Hamiltonian can be found in Table \ref{tab:parameters}. For the exchange constant $J_1, J_2$ and $J_3$ there are four interactions for each atomic site, and for $J_4$ there is just one.
\begin{figure}[t!]
\includegraphics{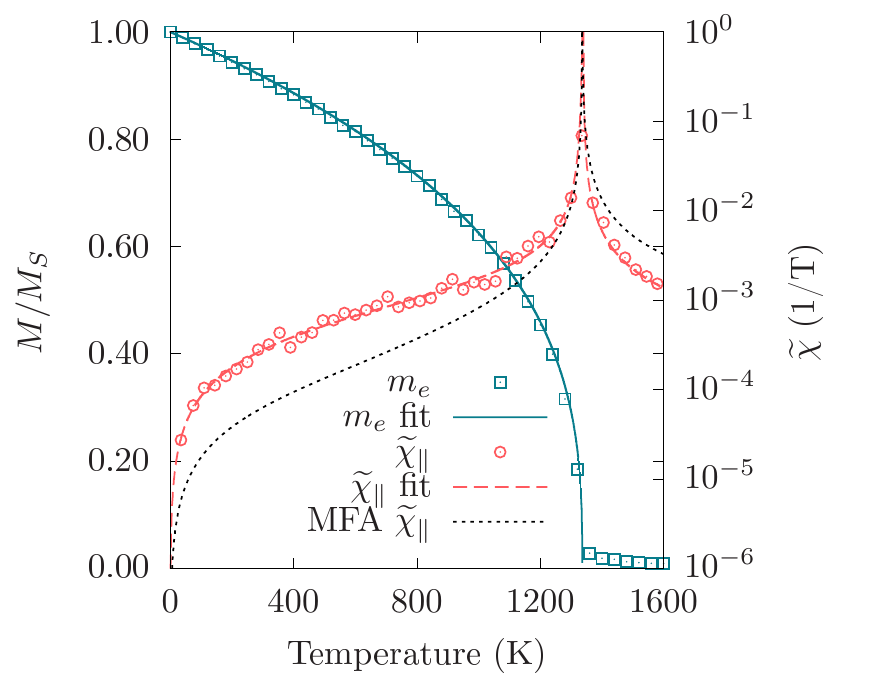}
\caption{The sublattice magnetization of \mnau as a function of temperature. The solid line is a fit to the expression $M/M_S=m_e(T) = (1 - T/T_N)^b$. The circular points show the longitudinal susceptibility, $\widetilde{\chi}_{\|}(T)$, calculated from ASD and the dashed line shows  $\widetilde{\chi}_{\|}(T)$ taken from MFA, which is discussed in Sec. \ref{sec:long}.}
\label{fig:mvt}
\end{figure}

\section{Transverse Dynamics of the Antiferromagnetic Landau-Lifshitz-Bloch Equation}
\begin{figure}[t!]
    \hspace*{-0.5cm}  
    \includegraphics{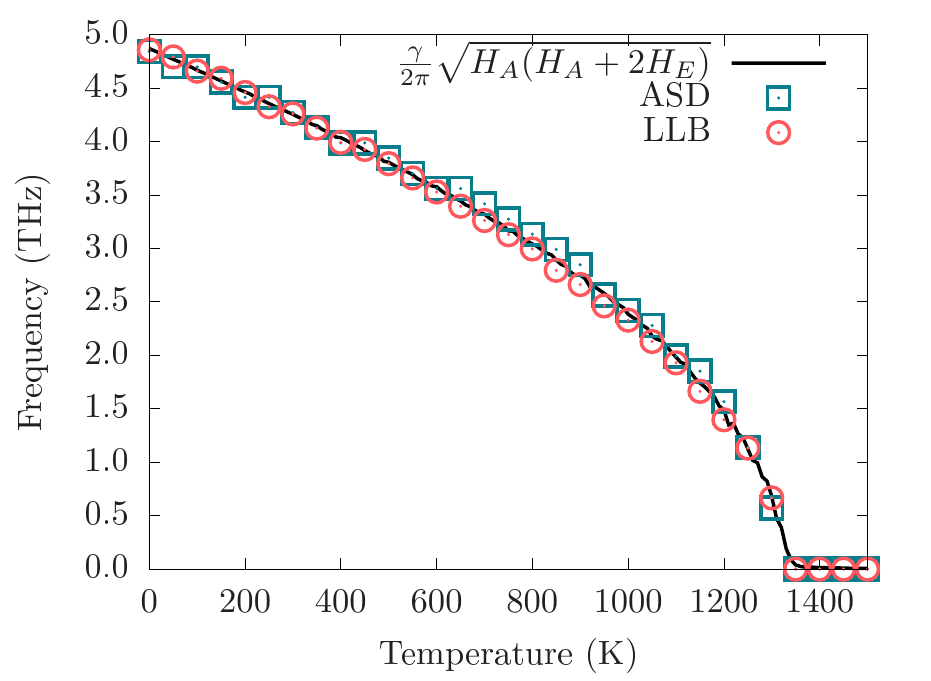}
	\caption{AFMR frequency of \mnau. The solid line is the result from Eq. \eqref{eq:afmr}, square points are from ASD, and circular points are the LLB simulations respectively.}
	\label{fig:afmr}
\end{figure}

The LLB model captures the  dynamics of a collection of exchange coupled atomistic spins, namely the dynamics of thermal averaged
$\mathbf{m}_i=\langle \mathbf{S}_i \rangle$. It was first derived by Garanin \cite{garanin} within a mean-field approximation for ferromagnets, which implies $\mathbf{m}_i=\mathbf{m}$ for lattice sites $i$. Extensions of the model for non-homogeneous magnetization states exist within micromagnetic computational approaches \cite{Kazantseva2008}.  More recently, the LLB model was extended to ferrimagnets by Atxitia et al \cite{Atxitia2012}. The LLB model contains both transverse and longitudinal relaxation terms and interpolates between Landau-Lifshitz-Gilbert equation at low temperatures (transverse dynamics) and the Bloch equation at high temperatures (longitudinal dynamics). 
To start with, a comparison of the transverse dynamics between the AFM-LLB and atomistic models is given. We begin with an LLB equation that neglects changes in the magnetization length, and purely describes the transverse dynamics,
\begin{equation}
\label{eq:llbperp}
    \frac{{d\mathbf{m}}_{\nu}}{dt}= - \gamma\left[\mathbf{m}_{\nu} \times \mathbf{H}_{\text {eff}, {\nu}}\right] - \gamma\alpha_{\perp} \frac{\left[\mathbf{m}_{\nu} \times\left[\mathbf{m}_{\nu} \times \mathbf{H}_{\mathrm{eff},{\nu}}\right]\right]}{m^{2}_{\nu}}
\end{equation}
The above looks remarkably similar to the LLG equation in that we have a precessional and relaxation term in both models.
In the above, $\mathbf{m}_\nu$ is the macrospin magnetization in sublattice $\nu$. It is not of constant length and its equilibrium value, $m_e(T)$ is temperature dependent. The dimensionless transverse damping parameter is given by,
\begin{equation}
  \alpha_{\perp} = \begin{cases} \lambda \Big(1-\frac{T}{3T_N}\Big), & T < T_N \\ \\
   \lambda \frac{2T}{3T_N}, & T > T_N
  \end{cases}
\end{equation}
\begin{figure}[t!]
    \hspace*{-0.5cm}  
    \includegraphics{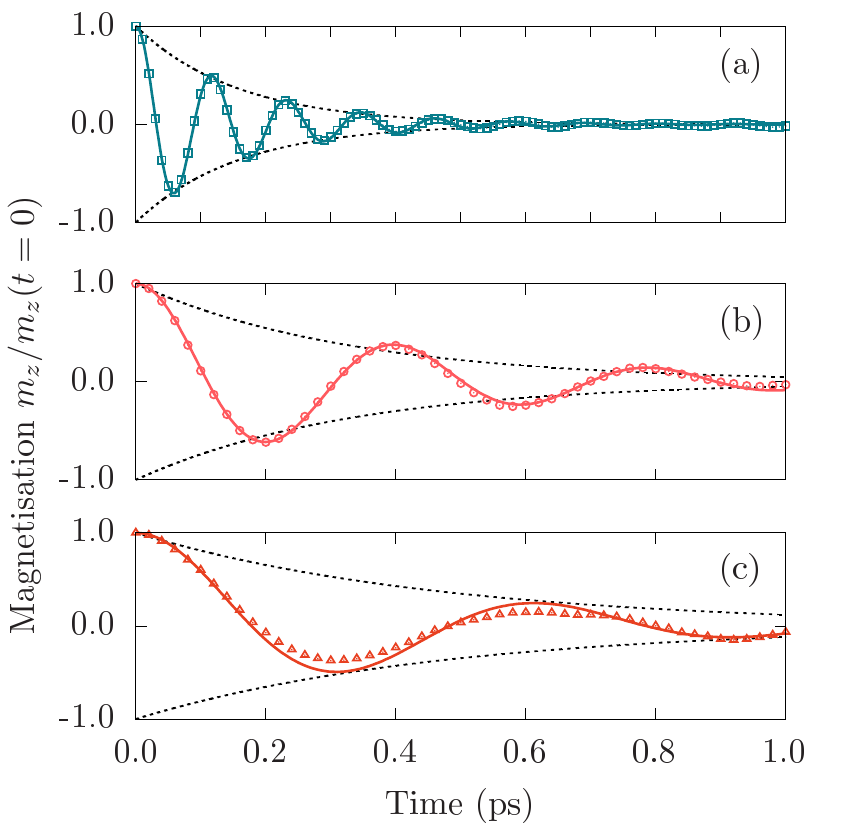}
	\caption{The $m_z$ motion following a rotation of of both sublattices by 20 degrees out of the easy plane at (a) $T = 300 $ K, (b) $T = 1000 $ K and (c) $T = 1200 $ K. Solid line are the AFM-LLB model and points are ASD. The dotted line shows the relaxation of the transverse dynamics. The relaxation time can be related to $\alpha_\perp$ through with Eq. \eqref{eq:trans_time}.}
	\label{fig:trans_comparison}
\end{figure}
It is worth noting that for situations where the magnetization length is fairly constant, $m_\nu = m_{e,\nu}$, Eq. \eqref{eq:llbperp} can be written in terms of a normalized vector, $\mathbf{n}_\nu = \mathbf{m}_\nu/m_{e,\nu}$.
\begin{equation}
\label{eq:llbperpnorm}
    \frac{{d\mathbf{n}}_{\nu}}{dt}= - \gamma\left[\mathbf{n}_{\nu} \times \mathbf{H}_{\text {eff}, {\nu}}\right] - \gamma\frac{\alpha_{\perp}}{m_{e,\nu}} 
    \left[\mathbf{n}_{\nu} \times\left[\mathbf{n}_{\nu} \times \mathbf{H}_{\mathrm{eff},{\nu}}\right]\right],
\end{equation}
which is identical to the LLG equation \cite{LLG1} with the identification $\alpha_{\rm{LLG}} = \alpha_{\perp} / m_e$. 
\\
\\
The \neel temperature, $T_N$, is taken from atomistic simulations of the equilibrium magnetization, $m_e(T)$. For these calculations, we simulate $30 \times 30 \times 30$ unit cells with periodic boundary conditions. The atomistic damping in Eq. \eqref{eq:sLLG}, $\lambda_i$, is set 1.0 for both sublattices to allow for a faster relaxation of the magnetization length. The equilibrium magnetization as a function of temperature can be found in Fig. \ref{fig:mvt}. The fit function, $m_e(T) = (1 - T/T_N)^b$ shown by the solid black line in Fig. \ref{fig:mvt}, has free fitting parameters $T_N$ and $b$ which provide a good estimate of the transition temperature for an infinite system. Fitting to the equilibrium magnetization from ASD yields $T_N = $ 1335 K and $b = 0.34$. A more accurate way to determine the critical temperature is through calculations of the parallel susceptibility, $\widetilde{\chi}_{\|}(T)$. In an atomistic spin model, the parallel susceptibility is found by measuring the fluctuations of the magnetization
\begin{equation}
\widetilde{\chi}_{\|}=\frac{\mu_s N}{k_{\rm{B}} T}\left(\left\langle S_{\|}^{2}\right\rangle-\left\langle S_{\|}\right\rangle^{2}\right)
\end{equation}
where $\langle ... \rangle$ is an ensemble average and $k_{\rm{B}} T$ is the thermal
energy. Due to the strong uniaxial easy-plane anisotropy in Mn\textsubscript{2}Au, the parallel component is taken as the average of the $x$ and $y$ components of the sublattice magnetization. It is worth noting that the thermal equilibrium parameters are the same for both sublattices. A value of $T_N = 1333$ K was extracted from fitting to $\widetilde{\chi}_{\|}(T)$ calculated from atomistic spin dynamics.
\\
\\
Moving away from atomistic spin dynamics and returning to the discussion of the AFM-LLB model -- for a single macrospin containing two sublattices with the magnetization length initialised at $m_{\nu} = m_e(T)$, the effective field, $\mathbf{H}_{\text {eff}, {\nu}}$, in Eq. \eqref{eq:llbperp} is given by 
\begin{equation}
\mathbf{H}_{\mathrm{eff}, \nu} = \mathbf{B}+\mathbf{H}_{a, \nu}+\frac{J_{0, \nu \kappa}}{\mu} \mathbf{\Pi}_{\kappa}
\label{eq:ferri_hamil}
\end{equation}
where $\mathbf{B}$ is the applied magnetic field, $\mathbf{H}_{a,\nu}$ is the anisotropy field, $J_{0, v \kappa}$ is the sum of the inter-lattice exchange (see Table \ref{tab:parameters}), and $\mathbf{\Pi}_{\kappa}$ is given by, $\mathbf{\Pi}_{\kappa}=-\left[\mathbf{m}_{\nu} \times\left[\mathbf{m}_{\nu} \times \mathbf{m}_{\kappa}\right]\right]/ m_{\nu}^{2}$ where $\mathbf{m}_\kappa$ is the magnetization vector of the second sublattice. The anisotropy field in the  macrospin model, $\mathbf{H}_{a,\nu}$, is defined as $\mathbf{H}_{a,\nu}  = \frac{2K(T)}{M_S}$ where the temperature dependence of the anisotropy constant is governed by Callen-Callen scaling;  $K(T) = K(0) m^3_e(T)$ \cite{CALLEN19661271}.
\begin{figure}[t!]
    \hspace*{-0.5cm}  
    \includegraphics{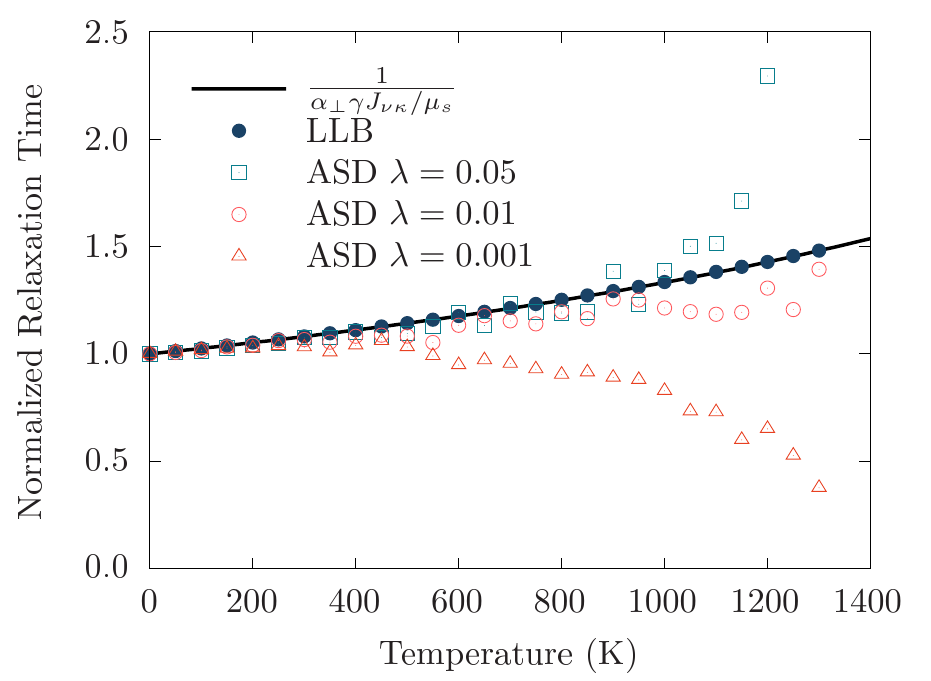}
	\caption{The transverse relaxation time, $\tau_\perp$ as function of temperature. Empty points represent ASD simulations with values of $\lambda = 0.05$ (square), $0.01$ (circle) and $0.001$ (triangles). Solid circles are the AFM-LLB model. The solid line is given by Eq. \eqref{eq:trans_time}.}
	\label{fig:trans_temp}
\end{figure}
\\
\\
To validate the temperature-dependent transverse dynamics described by the LLB model, we present simulation results showing the temperature dependence of the AFMR for \mnau using both atomistic s-LLG and AFM-LLB models, as well as a comparison to the Kittel equation for AFMR \cite{kittel1, kittel2}. Magnetic oscillations in AFMs have a resonance frequency in the THz range, orders of magnitude faster than the GHz range observed in ferromagnets. As a result of the complexity in current THz-signal generation methods \cite{superconductor-thz, qm-thz, linac-thz}, AFM oscillators have been proposed for use in adjustable room-temperature THz-frequency signal devices \cite{Checinski2017, PhysRevLett.116.207603, Khymyn2017}. While attempts to experimentally measure the in-plane AFMR in \mnau driven by \neel spin orbit torques have previously failed \cite{afmr_withdrawal}, it piqued the magnetism community's interest, highlighting the importance of not only measuring, but also calculating, the resonant frequency in AFMs. Since the AFMR sets the speed limit of the AFM dynamics, it is highly relevant for the development of ultrafast memory applications. For the atomistic simulations of the AFMR, the system is initially set to relax to its equilibrium magnetization for each given temperature. Once relaxed, all spins are rotated by 20 degrees out of the easy plane inducing a torque and thereby exciting the AFMR mode. The relaxation time of the transverse motion, $\tau_\perp$, can be related to the effective field and transverse damping via
\begin{equation}
    \label{eq:trans_time}
    \tau_\perp = \frac{m_e}{\alpha_\perp \omega_\mathrm{ex}}
\end{equation}
where $\omega_\mathrm{ex} = \gamma H_E= \gamma \frac{J_{0, \nu \kappa}}{\mu_s}m_e$. An example of the $m_z$ dynamics at $T = 0$ K can be seen in Fig. \ref{fig:trans_comparison}. The value of $\tau_{\perp}$ was found by fitting the sublattice magnetization
to the equation $m_z(t)/m_e =  \cos(\omega_{\rm{AFMR}}t) \exp(-t/\tau_{\perp})$.
\\
\\
A comparison of the temperature dependence of $\tau_\perp$ can be found in Fig. \ref{fig:trans_temp}. We calculate $\tau_\perp$ atomistically for values of $\lambda = 0.05, 0.01$ and $0.001$. Note for lower values of $\lambda$ we see a reduction in the relaxation time as we approach $T_N$. This effect is explained by the fact at lower damping we are further from equilibrium with the thermal bath and energy is predominantly transferred between the sublattices. The temperature dependence of $\tau_\perp$ in the LLB model is independent of $\lambda$. The exact form of the damping dependence on $\tau_{\perp}$ remains an open question.
\\
\\
The AFMR frequency of the dynamics was extracted via Fast Fourier Transform (FFT) of $m_z$ of a single sublattice. The results of this frequency analysis can be found in Fig. \ref{fig:afmr} showing excellent agreement between the models and the analytical expression. The analytic expression for the AFMR, shown by the solid line in Fig. \ref{fig:afmr}, is given by:
\begin{equation}
\label{eq:afmr}
f = \frac{\gamma}{2 \pi}\sqrt{H_A(H_A+2H_E)}
\end{equation}
where $H_A$ and $H_E$ are the anisotropy and exchange fields respectively.

\section{Longitudinal Relaxation}
\label{sec:long}
It has been demonstrated that a femtosecond laser-pulse can induce ultrafast magnetization dynamics in ferro- and ferri-magnets \cite{Beaurepaire,Radu2011,Ostler2012}. Whilst mechanisms leading to the ultrafast demagnetization in magnetic materials broadly are still under debate, in certain ferrimagnets, a single laser-pulse can induced switching of the magnetization \cite{Ostler2012}. The magnetic order dynamics in AFMs, i.e. \neel vector length dynamics, is less known.
In order to describe the magnetic order quenching in Mn$_2$Au, we introduce a term in the LLB equation that accounts for changes in the magnetization length. Here we are only interested in the longitudinal dynamics -- parallel to the magnetization direction -- and can therefore ignore the transverse terms.
\begin{equation}
\label{eq:llblong}
    \frac{d \mathbf{m}_{\nu}}{d t}=\gamma \alpha^{\mathrm{AF}}_{\|} \frac{\left(\mathbf{m}_{\nu} \cdot \mathbf{H}_{\mathrm{eff}, \nu}\right)}{m_{\nu}^{2}} \mathbf{m}_{\nu}
\end{equation}
The effective field for a purely longitudinal relaxation (where one can ignore any external, anisotropy or inter-macrospin exchange fields) is given by
\begin{equation}
\mathbf{H}_{\mathrm{eff}, \nu}^{\|}=\begin{cases}\frac{1}{2 \widetilde{\chi}_{\|}}\left(1-\frac{m^{2}_\nu}{m_{\mathrm{e}}^{2}}\right) \mathbf{m}_\nu, & T \lesssim T_{N} \\ \\
-\frac{1}{\widetilde{\chi}_{\|}}\left(1+\frac{3}{5} \frac{T_{N}}{T-T_{N}} m^{2}_\nu\right) \mathbf{m}_\nu, & T \gtrsim T_{\mathrm{c}} \end{cases}
\label{eq:ferro_hamil}
\end{equation}
where $\widetilde{\chi}_{\|}$ is the reduced longitudinal susceptibility. In this work we opt to use the susceptibility taken from a  MFA \cite{mfa}. We find that the agreement in the longitudinal relaxation between ASD and LLB when using $\widetilde{\chi}_{\|}$ taken from ASD simulations gave large discrepancies between the two models at lower temperature while using $\widetilde{\chi}_{\|}$  taken from a MFA gave excellent agreement for all temperatures. This problem has been noted elsewhere - Vogler et al mention that calculations of $\widetilde{\chi}_{\|}$ only work for hard magnetic materials with strong uniaxial anisotropy \cite{sus_problem2} with similar problems having been identified in ferrimagnets \cite{sus_problem1}. The reason for the difference is unclear, and should be investigated as part of future work. The method of calculating the MFA longitudinal susceptibility follows directly from \cite{Atxitia2017}. Using the same value for the exchange constants in ASD and MFA approaches yields a higher critical temperature compared to the atomistic results ($T_N^{\text{MFA}} > T_N^{\text{ASD}}$). Therefore, we rescale the exchange constant to match the value for the \neel temperature extracted from Fig. \ref{fig:mvt}. The process of using a scaled MFA susceptibility has been seen previously in atomistic modelling of GdFeCo \cite{PhysRevB.84.024407}.
\begin{figure}
\includegraphics{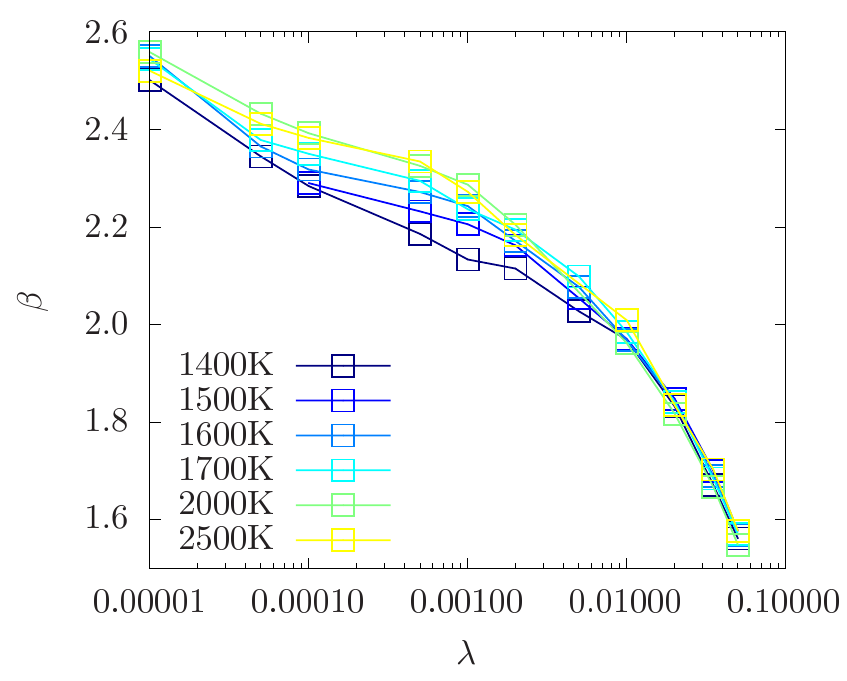}
\caption{The damping dependence of $\beta$. The MFA LLB equation \cite {garanin} is fitted to the longitudinal dynamics from atomistic simulations following a step change in temperature from $T = 0$ K to $T > T_N$. All other parameters in the LLB besides $\beta$ remain fixed.}
\label{fig:beta}
\end{figure}
Returning to the discussion of the longitudinal relaxation in the LLB equation, assuming that the 
macrospin magnetization remains along single axis, $\mathbf{m}_\nu = [m_\nu, 0, 0]$, Eq. \eqref{eq:llblong} simplifies to
\begin{equation}
    \frac{dm_{\nu}}{dt} = \gamma \alpha^{\mathrm{AF}}_{\|} H^{\|}_{\mathrm{eff}, \nu}.
\end{equation}
Experimental works studying the ultrafast magnetic order dynamics in the FM and AFM phases of Dy have shown that the AFM phase exhibits faster dynamics than the FM phase \cite{PhysRevLett.119.197202}. This speed up is attributed to the exchange of angular momentum between sublattices. For \mnau, it will be difficult to conduct a similar comparative study due to the absence of a FM phase. However, a recent theoretical work by Jakobs \& Atxitia \cite{florian} has concluded that the speed up of the AFM dynamics comes from the exchange-enhancement of the effective damping parameter
\begin{equation}
\label{eq:afdamp}
    \alpha^{\mathrm{AF}}_{\|} =\alpha^{\mathrm{FM}}_{\|}\left(1+\frac{2}{z\left|m_{\nu}\right|^\beta}\right)
\end{equation}
where, $\alpha^{\mathrm{FM}}_{\|}$ is the ferromagnetic longitudinal damping parameter (where the relaxation is solely due to the dissipation of angular momentum to the heat bath) and is defined by
\begin{equation}
\alpha^{\mathrm{FM}}_{\|} = \lambda \frac{2 T}{3 T_N}
\label{eq:ef_damp}
\end{equation}
In Eq. \eqref{eq:afdamp}, theory predicts that $z$ is the number of nearest neighbours antiferromagnetically coupled to a given spin. While for a simple cubic with only nearest neighbours this number is 6, for \mnau this would correspond to 5, as sketched in Fig. \ref{fig:mnau_cell}.
Our simulations show that $z \approx 6.0$. The exponent $\beta$ in Eq. \eqref{eq:afdamp} is a phenomenological parameter, necessary for the description of high non-equilibrium situations where the temperature goes well above the critical temperature. While for small deviations from the equilibrium of the magnetic order parameter, $\beta=1$, for larger deviations, for example when the temperature of the system changes from $T=0$ K to $T=2T_N$, the exponent takes a value of around 2.
We also find that alongside the temperature dependence of the $\beta$ exponent, it also exhibits a damping dependence, which can be found in Fig. \ref{fig:beta}. To determine the damping dependence of $\beta$, numeric solutions of the LLB equation after step change in temperature from $T = 0$ K to $T > T_N$ were fitted against the corresponding atomistic simulations. For the fitting, we use a MFA form of the effective field that is not dependent on any atomistically derived parameters. Instead of using Eq. \eqref{eq:ferro_hamil}, the effective field is given by \cite{Atxitia2012}
\begin{equation}
H_{\mathrm{eff}, \nu}^{\|}=\frac{\left(m_{\nu}-m_{0, \nu}\right)}{\mu_{s} \beta L^{\prime}\left(\xi_{\nu}\right)}
\end{equation}
in the above, $m_{0, \nu}$ is not the equilibrium magnetization and is given by $m_{0, \nu} = L\left(\xi_{\nu}\right)$ where $L\left(\xi_{\nu}\right) = \operatorname{coth}(\xi_{\nu})-1 / \xi_{\nu}$ is the Langevin function and $L^{\prime}(\xi_{\nu})=d L / d \xi_{\nu}$ with 
$\xi_{\nu}=\beta \mu_{s} H_{\nu}^{\mathrm{MFA}}$ and $\beta=1 / k_{\mathrm{B}} T$.
The MFA field is defined as $H^{\mathrm{MFA}}_{\nu} = 3 k_B T_{N} m_{\kappa}$ \cite{garanin}.
\begin{figure}
\includegraphics{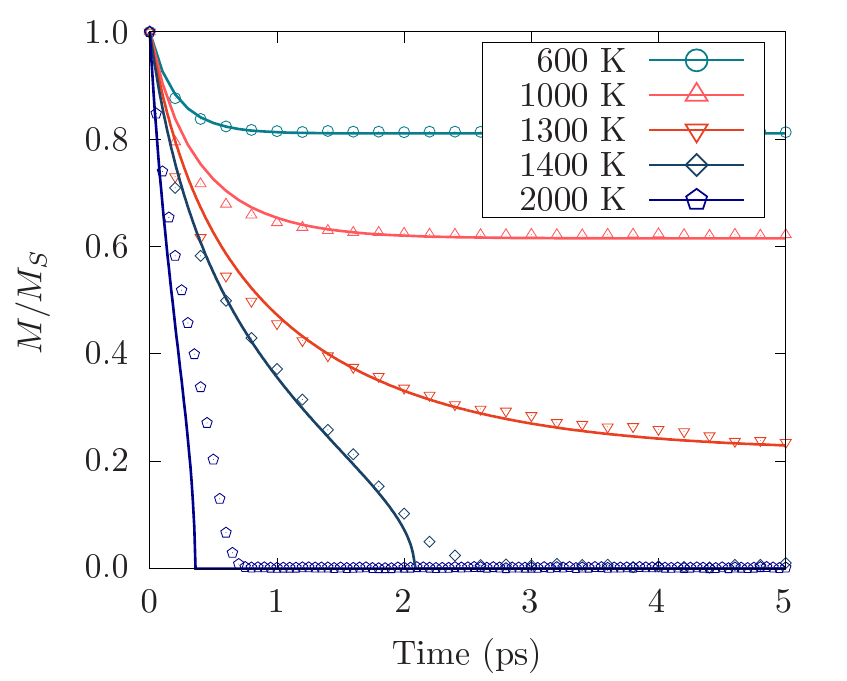}
\caption{A comparison at five different temperatures between the AFM-LLB and the atomistic modelling. Symbols represent the atomistic spin dynamics results and the solid lines correspond to the AFM-LLB. No stochastic noise was present in the AFM-LLB simulations.}
\label{fig:long}
\end{figure}
\\
\\
Using the results of our determination of $\beta$ for a value of $\lambda=0.01$ we have furthermore carried out simulations of the longitudinal dynamics after a temperature step (Fig.~\ref{fig:long}). In these temperature step simulations, we begin with a perfectly ordered AFM configuration of \mnau at $T = 0$ K, we then apply a temperature step and record the relaxation to the equilibrium magnetization. Fig. \ref{fig:long} shows the longitudinal relaxation for 5 different temperatures (with $\lambda=0.01$ for all simulations), the points are simulations using the stochastic LLG equation and the solid lines are from the LLB model. The agreement at lower temperatures is excellent. At temperatures well above $T_N$, finite size effects lead to a longer relaxation time in the atomistic model. It is also worth stressing that instantaneous step changes in temperature from 0 Kelvin to near $T_N$ are radical, and some disagreement would be expected.

\begin{figure}
\includegraphics{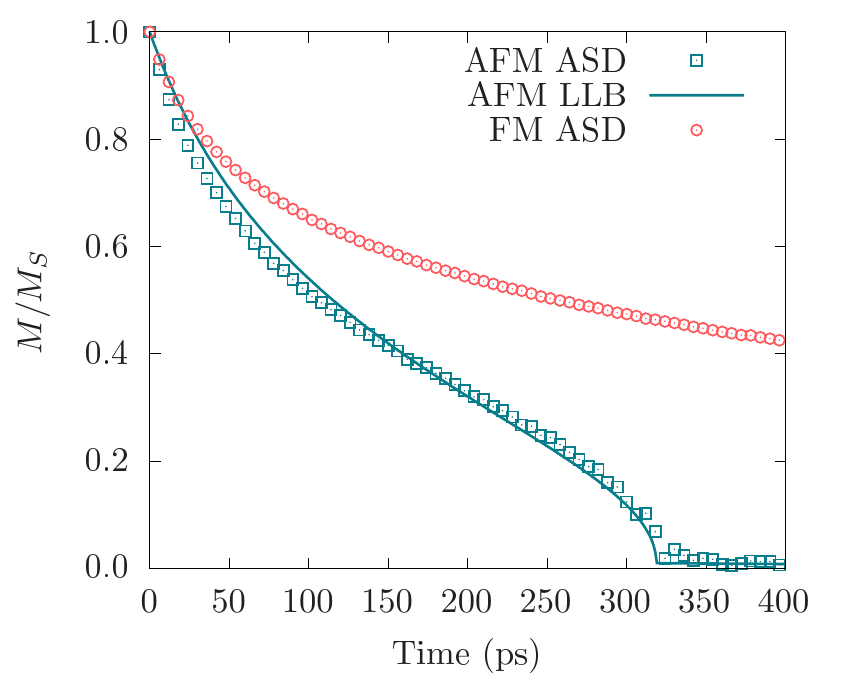}
\caption{The longitudinal relaxation following a step change in temperature from 0 K to 1400 K. The Gilbert damping was set to 0.00005. Square points are from ASD simulations of the ground state AFM configuration of mn\textsubscript{2}Au. Circular points are from ASD simulations representing an FM configuration of \mnau where the absolute values of $J_1$ to $J_4$ have been used. The solid line shows the AFM-LLB.}
\label{fig:low_damp}
\end{figure}

\begin{figure}
\includegraphics{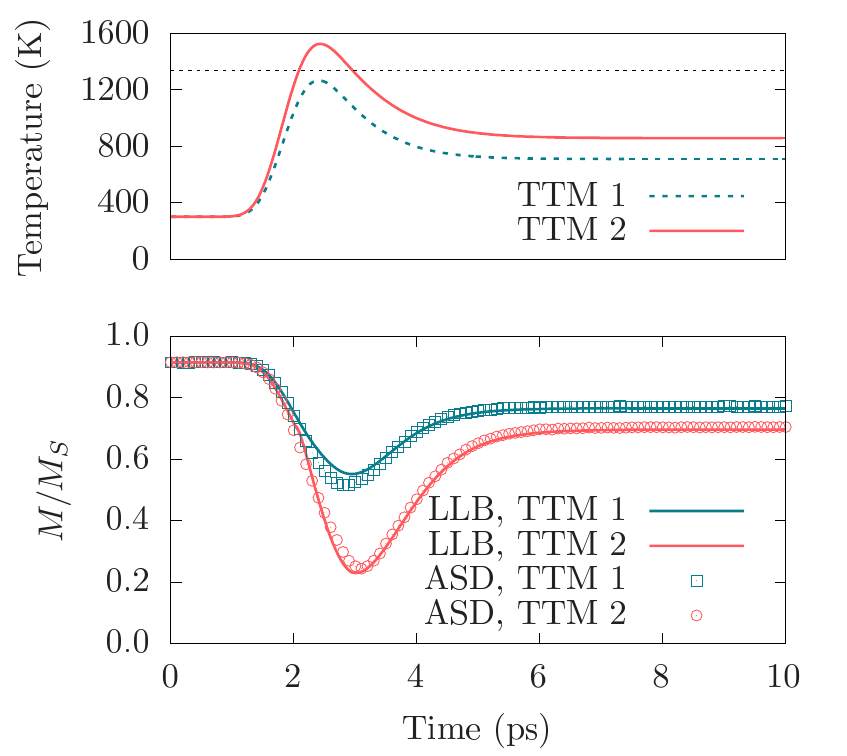}
\caption{(Top) The electronic temperature dynamics using a TTM. Dotted line shows the \neel temperature. (Bottom) Magnetization dynamics from atomistic simulations (black squares) and the LLB equation using a ferromagnetic (dotted lines) and ferrimagnetic (solid lines).}
\label{fig:ttm}
\end{figure}

To highlight the importance of the $\beta$ exponent in Eq. \eqref{eq:afdamp}, we performed ASD and LLB simulations with values for $\lambda = 0.00005$ for a step change to above the critical temperature, as shown in Fig. \ref{fig:low_damp}. We also include ASD results for an entirely ferromagnetic exchange to show the difference in the relaxation rates between FM and AFM.
We find that relaxation dynamics in AFMs is defined by two distinct processes, (i) a exponential decay due to dissipation of angular momentum to the heat-bath and (ii) a power-law decay due to angular momentum exchange between sublattices.
 While for the FM version of \mnau the relaxation dynamics is described by an exponential decay, for AFMs, the exponential decay only dominates for values of the order parameter $n > 1/3$.  As the magnetic order reduces to small values, the rate of angular momentum dissipation remains constant, $(\mu_{\rm{at}}/\gamma)\dot{n} \approx (2/3)  \lambda  k_B  T/n^{\beta-1} $,  leading to a power-law decay.
\\
\\
Although in the simulations a step-like increase of temperature is easy to achieve, experimentally, a strong and rapid heating of the system is possible by applying, for example, a femtosecond laser pulse. In these scenarios the temperature of the system is difficult to determine, however, given semi-classical considerations \cite{Chen2006} one can define a temperature of an electron and a phonon bath. A laser pulse will couple more strongly to the electron system giving large and rapid temperature increases, in hundreds of femtoseconds, to a temperature above $T_N$, however, the electron system also quickly cools downs on the picosecond time scale by transferring energy to the phonon system via electron-phonon coupling. The electronic temperature is calculated using a two-temperature model (TTM) \cite{ttm}.
Fig. \ref{fig:ttm} shows the longitudinal magnetization dynamics for a transient change in temperature following heating from a laser pulse. We begin at $T = 300$ K then heat to just below and above $T_N$, shown as TTM 1 and TTM 2 in Fig. \ref{fig:ttm}.  What is also worth noting is that the AFM-LLB accurately captures both remagnetization and demagnetization processes. 
\\
\\
These results show that the exchange-enhanced damping derived by Jakobs \& Atxitia is successful in describing the faster relaxation times within the LLB model, especially at lower temperatures. It also shows good agreement when using a TTM, which is highly applicable to simulations of all-optical switching in AFMs. 

\section{Domain Wall Motion}
\begin{figure}[t!]
\includegraphics{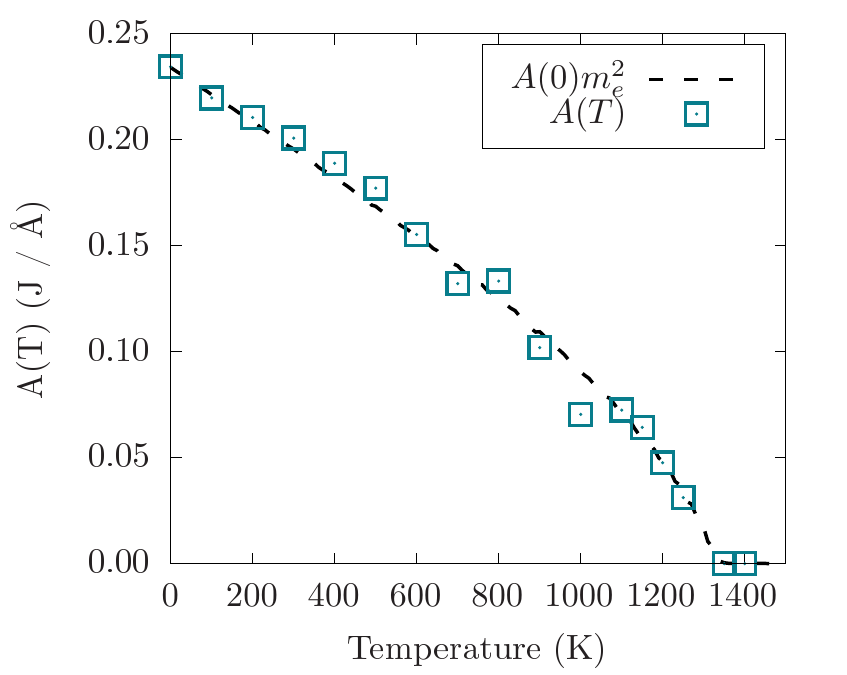}
\caption{Exchange stiffness as a function of temperature from atomistic simulations. Dotted line shows the proportionality to the equilibrium magnetization with $A(0)$ taken from Eq. \eqref{eq:A0}.}
\label{fig::A(T)}
\end{figure}

\begin{figure}[t!]
    \includegraphics{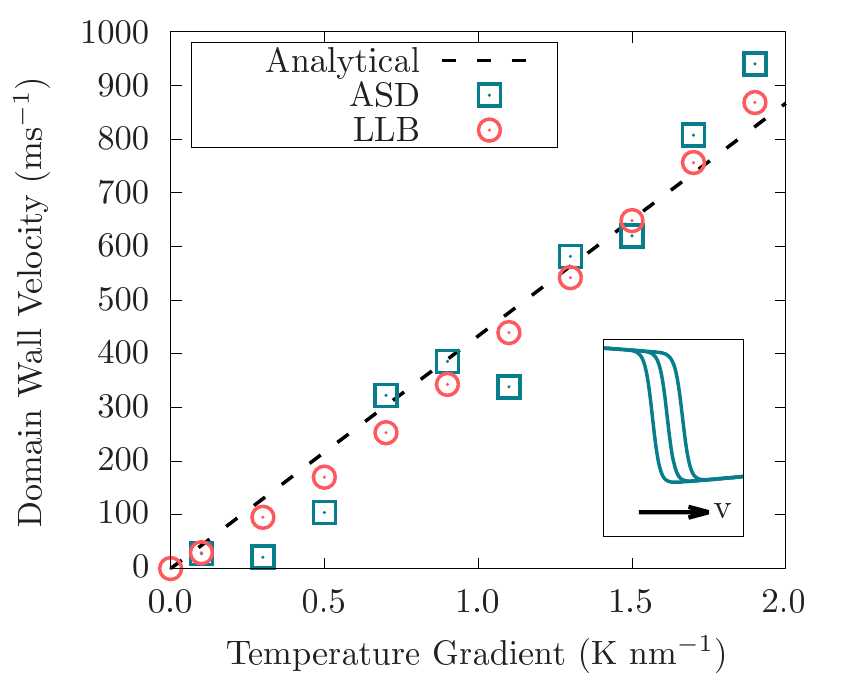}
	\caption{Domain Wall velocity as a function of the temperature gradient. $\lambda$ was set to 0.01 for the simulations. The solid line is given by Eq. \eqref{eq:dw_vel}. The inset shows an example of the domain wall at three different timesteps. The wall is moving from left to right in the figure.}
	\label{fig:dw}
\end{figure}

    The manipulation of domain walls (DW) have been proposed for use in the next generation of logic and memory devices \cite{doi:10.1126/science.1108813, doi:10.1126/science.1145799}. The motion of the DWs can be modulated using spinwaves \cite{dw_spinwaves1,dw_spinwaves2}, spin currents \cite{dw_currents1,dw_current2,dw_current3} or external magnetic fields \cite{walker,Shibata2011}, for example. Another possibility is to drive DW motion by thermal gradients \cite{dw_thermal1,dw_thermal2}. Under thermal gradient, motion of the domain wall is induced towards the hotter end of the nanowire due to a reduction in the free energy, $\Delta F(T)$. For large thermal gradients in ferromagnets, there is a precession of the internal magnetization, which leads to a reduction in the domain wall velocity, known as the Walker breakdown \cite{walker}. In previous ASD simulations of AFMs, it has been shown that there is no breakdown in the DW velocity due to symmetry of the torques acting on the DW \cite{unai_dw}, meaning DWs in AFMs are able to reach higher speeds.
    \\
    \\
    Conventional micromagnetic methods have limitations in the  modelling of thermally driven DWs, as they cannot account for the dynamic changes in the magnetization length. It has been shown experimentally that ultrashort laser pulses are able to drive DW motion \cite{dw_thermal2}. In these scenarios, heating and cooling will play a role, and the effects of this cannot be captured using an LLG based micromagnetic model. This could be simulated using an atomistic approach, but the calculations are expensive and require large ensembles to reduce statistical noise. The LLB model overcomes these issues, making it an important tool for temperature dependant calculations of topological structures such as domain-wall, spin-spirals and skyrmions. As an example of simulations that require both accurate descriptions of the longitudinal and transverse dynamics, we present a comparison of the DW motion under a thermal gradient using both atomistic and AFM-LLB models and compare to previously derived analytical expressions for the DW velocity. As we are considering both longitudinal and transverse dynamics, the LLB equation now reads
\begin{eqnarray}
\label{llbafm}
\frac{{d\mathbf{m}}_{\nu}}{dt}= \gamma\left[\mathbf{m}_{\nu} \times \mathbf{H}_{\text {eff}, \nu}\right]-\gamma \alpha_{\|}^{\nu} \frac{\left(\mathbf{m}_{\nu} \cdot \mathbf{H}_{\text {eff}, \nu}\right)}{m_{\nu}^{2}} \mathbf{m}_{\nu} \nonumber \\  
- \gamma \alpha_{\perp}^{\nu} \frac{\left[\mathbf{m}_{\nu} \times\left[\mathbf{m}_{\nu} \times \mathbf{H}_{\text {eff}, \nu}\right]\right]}{m_{\nu}^{2}}
\end{eqnarray}
To be able to describe DWs within an LLB framework, we introduce a term into the LLB Hamiltonian that describes the exchange coupling between neighbouring macrospins that is given by:
\begin{equation}
    \mathbf{H}_{\mathrm{ex}}^{v,i}=\frac{2 A(T)}{d^{2} M_{S} m_{e}^{2}} \sum_{j}\left(\mathbf{m}_{v, j} - \mathbf{m}_{v, i}\right)
\end{equation}
Where $A(T)$ is the exchange stiffness, $d^2$ is the surface area between neighbouring macrospins, and $M_S$ is the saturation magnetization. The value and temperature dependence of the exchange stiffness is calculated using atomistic simulations of the DW width, $\delta_0$. The relation that links the DW width to the exchange stiffness is given by
\begin{equation}
\delta_0(T) =\sqrt{\frac{A(T)}{K(T)}}
\end{equation}
where $K(T)$ is the anisotropy energy where again we use Callen-Callen scaling \cite{CALLEN19661271}, $K(T) = K m^3_e(T)$. The exchange stiffness, $A(T)$ scales with $m_e^2$ within a MFA. Fig. \ref{fig::A(T)} shows the exchange stiffness as a function of temperature with points representing atomistic simulation results, and the dotted line showing the scaling law $m_e^2$.
\\
\\
For the atomistic simulations of the domain wall width, we ensure the length of the system in the $x$-direction is much wider than the zero Kelvin domain wall width. The Mn sites at the left hand boundary of the slab remain fixed in an anti-parallel alignment to the spins at the right hand boundary. The remainder of the spins are allowed to relax to form a domain wall profile. The reduced magnetization, $m(x)$, is then fitted to a hyperbolic tangent function to find the width parameter:
\begin{figure}[t!]
\includegraphics{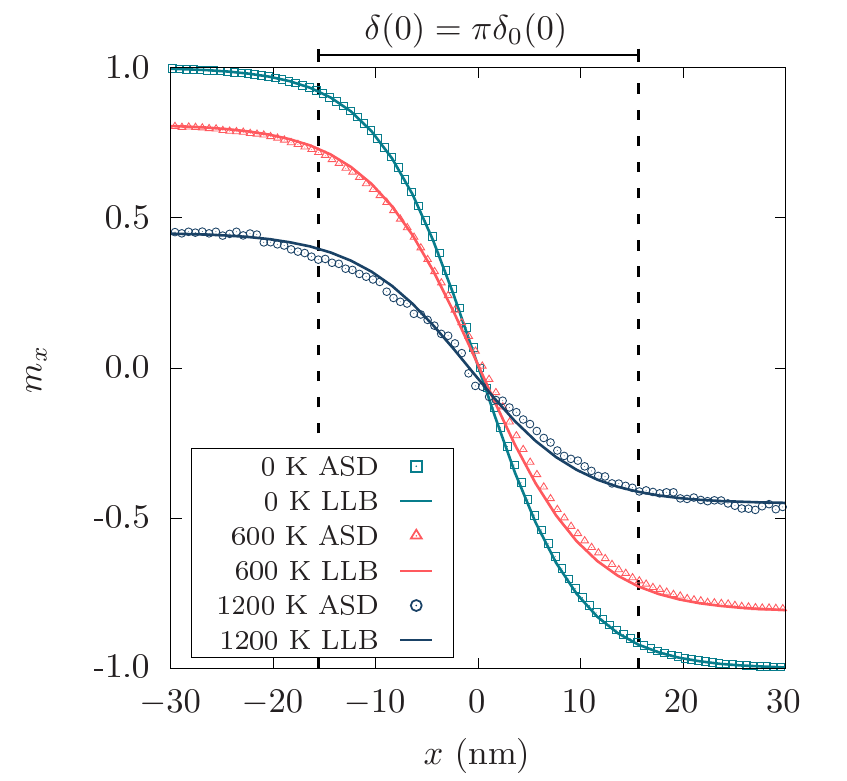}
\caption{The domain wall width at 0 K, 600 K and 1200 K. Solid lines are from LLB simulations and points are from ASD. The \neel domain-wall width is given by $\delta(T) = \pi \delta_0(T)$. The dotted lines show the width of the domain wall at 0 K.}
\label{fig:dw_comparison}
\end{figure}
\begin{equation}
    m(x) = m_e \tanh \Big( \frac{x-b}{\delta_0} \Big)
    \label{eq:dw_prof}
\end{equation}
where $b$ is the central position of the DW, and $\delta_0$ is the wall-width parameter.
A comparison of the DW profile at $T = $0 K can be found in Fig. \ref{fig:dw_comparison}. The \neel wall width can be related to the wall width parameter through $\delta = \pi \delta_0$. We calculate $\delta(0) = 31.2$ nm from atomistic simulations and see good agreement between ASD and LLB in Fig. \ref{fig:dw_comparison} for intermediate and high temperatures.
\\
\\
It is possible to relate the zero temperature exchange stiffness $A(0)$ directly to the $J_{ij}$ values and lattice constants found in Table \ref{tab:parameters} through the equation \cite{Atxitia2010}
\begin{equation}
    A(0)=\frac{1}{V_{0}} \sum_{i,v} \frac{\left|J_{i,v} \right|}{\left(a_{i,v}\right)^{2}}
    \label{eq:A0}
\end{equation}
where $V_0$ is the unit cell volume, $v$ represents all the neighbours for site $i$ where $x$ or $y$ are non-zero as the domain wall forms in the $xy$-plane, and $a_{v}$ is the absolute distance between site $i$ and site $v$. In total, there are 12 interactions that contribute to $A(0)$. Referring to Fig. \ref{fig:mnau_cell}, the only interaction that doesn't contribute is $J_4$. Using the values for $J_1$ to $J_3$ in Table \ref{tab:parameters} yields $A(T = 0 \ \mathrm{K}) = 2.34 \times 10^{-11} $ J/m, which agrees well with calculations of the exchange stiffness through atomistic simulations of the DW width, as shown in Fig. \ref{fig::A(T)} 
We should emphasize here that we use a different equation for the exchange coupling to Chen et al \cite{chen_afm_llb}. They use an exchange term that sums over the opposite sublattice in the neighbouring macrospins. This term would be non-zero when sublattices in neighbouring macrospins are parallel, leading to a disagreement between LLB and ASD simulations of the AFMR (see Sec. II).
\\
\\
For the atomistic calculations of the DW velocity, we begin with a DW initialised at 0 Kelvin along the $x-$direction, we then apply a linear thermal gradient in the direction of the DW. 
It has been shown that the DW velocity induced by a thermal gradient in an AFM can be approximated by \cite{Schlickeiser_dw}
\begin{equation}
\label{eq:dw_vel}
v_{\mathrm{DW}}^{\mathrm{LLB}}=\frac{2 \gamma }{M_{s} \alpha_\perp} \frac{\partial T}{\partial z} \frac{\partial A}{\partial T}
\end{equation}
As seen in previous analyses of domain walls \cite{Schlickeiser_dw, unai_dw}, the weak temperature dependence of $\alpha_\perp$ is neglected and the exchange stiffness is  linearized to $dA / dT \approx A(0)/T_C$. Fig. \ref{fig:dw} shows the domain wall velocity for \mnau in atomistic and LLB simulations due to a thermal gradient, showing good agreement with Eq. \eqref{eq:dw_vel}. The velocity is calculated by, once again, fitting the DW profile to Eq. \eqref{eq:dw_prof} to find the centre of the domain wall, then tracking the movement of this central position in time. While the relation appears linear in Fig. \ref{fig:dw}, in reality the velocity will reach a saturation point governed by the magnon group velocity \cite{PhysRevLett.117.017202}. In the atomistic modelling of thermally induced DW motion, Selzer et al \cite{unai_dw} state that there is no acceleration phase, and the DW moves with constant velocity, which is what we observe in \mnau in both LLG and LLB simulations.

\section{Summary}
 Using a multiscale approach, it has been demonstrated that antiferromagnets such as \mnau can be modelled micromagnetically on micrometer length scales using an AFM-LLB model. We started with parameters from previous work for the exchange, anisotropy and magnetic moment which were fed into an atomistic model of Mn$_2$Au. Then, we calculated the temperature dependent parameters $m_e(T)$, $A(T)$, $\widetilde{\chi}_{\|}(T)$ using atomistic spin dynamics before using these as input into the AFM-LLB. 
 \\
 \\
To verify the dynamics described by the proposed LLB model for Mn$_2$Au, first we presented consistent results between ASD, LLB and the Kittel relation for the antiferromagnetic resonance frequency. As well as serving as a validation of the transverse dynamics, we have revealed the first estimate of the in-plane AFMR frequency in Mn\textsubscript{2}Au. The fact the resonant frequency sits within the THz range opens the possibility sub-picosecond switching and generation of THz electromagnetic signals at ambient temperatures. Second, we compared ASD simulations results to the  AFM-LLB model for the longitudinal dynamics following step changes in temperature and laser pulse heating - both of which give excellent agreement between the models.  Finally, we provided a comparison of the domain wall motion due to a thermal gradient as an example of how the LLB model for an AFM can be utilised on $\mu$m length and $\mu$s timescales - opening the door to micromagnetic simulations of AFM materials for use in realistic spintronic devices. While questions such as the damping dependence of the transverse relaxation time, and the poor agreement for the longitudinal relaxation between ASD and LLB with $\widetilde{\chi}$ from ASD remain unanswered, the AFM-LLB opens the possibility for the description laser induced local thermal gradients on length-scales and the benchmark of other thermodynamic effects in AFMs on micrometer length-scales.

\section{Acknowledgements}
This work was supported by the EPSRC TERASWITCH project (project ID EP/T027916/1). Simulations were undertaken using the HPC cluster at Sheffield Hallam University and resources provided by the Cambridge Tier-2 system operated by the University of Cambridge Research Computing Service (www.hpc.cam.ac.uk) funded by EPSRC Tier-2 capital grant EP/P020259/1.
U. A. gratefully acknowledge support by the Deutsche Forschungsgemeinschaft (DFG, German Research Foundation)—Project-ID 328545488—TRR 227, Project No. A08. J. H. and U.A. acknowledge support from COST-Action CA17123.
\newpage
\bibliography{ref}
\end{document}